\documentclass[aps,prb,twocolumn,superscriptaddress,amsmath,amsfonts]{revtex4-1}

\usepackage{graphicx}
\usepackage{dcolumn}
\usepackage{bm}
\usepackage{graphicx}
\usepackage{comment}
\usepackage{braket}
\usepackage{txfonts}
\usepackage{physics}
\usepackage{color}
\newcommand{\nn}{\nonumber}
\usepackage{times}
\begin{document}

\title{Dissipationless Spin Current Generation in Kitaev's Chiral Spin Liquid } 

\author{Daichi Takikawa}
\email{takikawa@blade.mp.es.osaka-u.ac.jp}
\affiliation{Department of Materials Engineering Science, Osaka University, Toyonaka 560-8531, Japan}
\author{Masahiko G. Yamada}
\affiliation{Department of Materials Engineering Science, Osaka University, Toyonaka 560-8531, Japan}
\author{Satoshi Fujimoto}
\affiliation{Department of Materials Engineering Science, Osaka University, Toyonaka 560-8531, Japan}

\date{\today}

\begin{abstract}
$\alpha$-RuCl$_3$ is a promising candidate material for the Kitaev spin liquid state where the half quantization of the thermal Hall effect, suggesting a topological character, has been observed.
Here we propose a more direct signature of a chiral Majorana edge mode which emerges in a universal scaling of the Drude weight of the edge spin Seebeck effect in the Kitaev model.
Moreover, the absence of backscatterings in the chiral edge mode results in the generation of a dissipationless spin current in spite of an extremely short spin correlation length close to a lattice constant in the bulk.
This result is not only experimentally observable, but also opens a way towards spintronics application of Kitaev materials.
\end{abstract}

\maketitle

{\it Introduction.}--- Quantum spin liquids have been the focus of attention for many years, both theoretically and experimentally. 
They do not have long-range magnetic order despite their strong spin correlation.
Recently, it has been pointed out that the spin liquids are suitable for nano-scale spintronics devices~\cite{jauho1994,berger1996,sinova2004,kato2004,wunderlich2005,azevedo2005,saitoh2006,valenzuela2006,adachi2011,flipse2014,daimon2016,hirobe2017}. 
While quantum fluctuations rather stabilize the exotic states even in atomic scale,
fractionalized excitations characterizing spin liquids can indeed carry spin currents~\cite{hirobe2017}. 
On the other hand, the Kitaev spin liquid described by
the two-dimensional (2D) Kitaev model is potentially realized in 2D honeycomb materials such as $\alpha$-RuCl$_3$~\cite{kitaev2006,baskaran2007,jackeli2009,liu2011long,ye2012direct,rau2014,knolle2014,hermanns2014,plumb2014,yamaji2014,kubota2015successive,nasu2015,hermanns2015,johnson2015monoclinic,RuCl3_exp1,cao2016low,williams2016incommensurate,song2016low,baek2017evidence,yamada2017mof,nasu2017thermal,yamada2017xsl,zheng2017gapless,hentrich2018unusual,janvsa2018observation,zhu2018robust,vinkler2018approximately,ye2018quantization,udagawa2018vison,rusnavcko2019kitaev,gordon2019theory,zhang2019vison,nasu2019nonequilibrium,yamada2020,yamada2020electric,yamada2021,takahashi2021}. 
The Kitaev materials may provide another possible root to nano-scale spintronics application. 
The Kitaev model is a spin system in which Ising-like exchange interactions depending on bond directions act on $S = 1/2$ spins localized on each site of the honeycomb lattice. 
In terms of the Majorana fermion representation, the model
is exactly solvable, and exhibits a quantum spin liquid state regardless of the system size~\cite{kitaev2006}.
When time-reversal symmetry is broken by an applied magnetic field, the system becomes a chiral spin liquid state with a chiral Majorana edge, resulting in the Ising topological order in the bulk.
The observation of the half-quantized thermal Hall effect, which is a signature of  chiral Majorana fermions, has been reported~\cite{kasahara2018}.
Although further experimental observations which support the half-quantization of the thermal Hall effect in $\alpha$-RuCl$_3$ are accumulating~\cite{kasahara2018,yokoi2020,yamashita2020},
it is still an important issue to confirm the existence of
Majorana fermions from different and more direct approaches.

Here we propose the universal scaling of the Drude weight of the edge spin Seebeck effect of the Kitaev model as a definitive evidence of the existence of Majorana fermions.
The spin Seebeck effect is a phenomenon in which 
temperature difference imposed on opposite sides of a sample produces the flow of a spin current.
In the Kitaev spin liquid, spins are fractionalized into Majorana fermions, and it has an extremely short spin correlation length close to a lattice constant in the bulk. In spite of this feature, 
it is found that a dissipationless spin current flow at the edge of the system, which leads to
the spin Seebeck effect.
We here use the temperature dependence of the Drude weight of this edge spin Seebeck effect as the signature of the Majorana edge mode.
This signature is expected to be more general and stable than the thermal Hall effect.
This is because that it does not require
quantitatively precise measurements like
the half-quantized thermal Hall effect, and the qualitative temperature dependence is not expected to be affected even if it is disturbed by other degrees of freedom, such as phonons. 

Furthermore, our proposal is also useful for
the application to spintronics devices.
The Kitaev spin liquid state remains stable down to atomic scale, and enables the generation of a spin current without dissipation due to the existence of a chiral Majorana edge mode.
 The spin current may be detected via the measurement of the
 surface magnetization generated by the spin accumulation, or the inverse spin Hall effect.
This makes it possible to fabricate a highly integrated device with substantial efficiency to generate a spin current, paving the way to \emph{Kitaev spintronics}.

\begin{figure}[b]
\includegraphics[width=8cm]{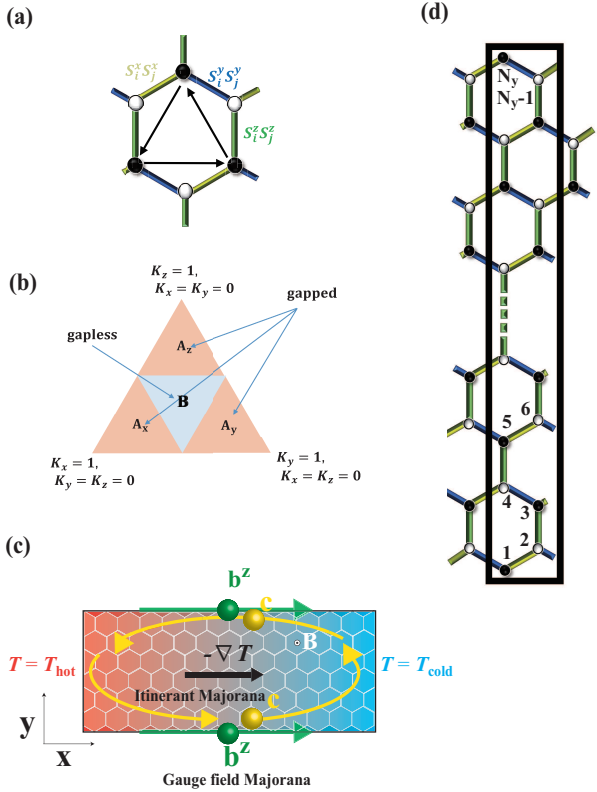} 
\caption{({a}) Structure of the Kitaev model. Yellow, blue, and green bonds represent $x$-, $y$-, and $z$-bonds, respectively. Black arrows represent the sign convention of the next-nearest-neighbor hoppings.
({b}) Phase diagram of the Kitaev model. The orange region is the gapped $A$-phase, while the blue region is the gapless $B$-phase. ({c}) Experimental set-up of the spin Seebeck effect. ({d}) Unit cell of the system with open edges in the $y$ direction.}
\label{fig:setup}
\end{figure}

{\it Model set-up.}---Our model set-up for the edge spin Zeebeck effect is
illustrated in 
{Fig.~\ref{fig:setup}}(c).
We consider an open boundary condition for the $y$-direction and a periodic boundary condition for the $x$-direction.
The unit cell of the system is shown in {Fig.~\ref{fig:setup}}(d).
We label unit cells as $l = 1,\dots,L_x$, where $L_x$ is the number of unit cells, from left to right.
For this configuration, the gauge-field Majorana fermions 
included in the top and bottom spins at open edges, $b^z_{l,1}$ and
$b^z_{l,N}$, can not form the $Z_2$ gauge fields~\cite{de2018,mizoguchi2019,mizoguchi2020,minakawa2020}.
Thus, the perturbative calculations within the vortex-free sector described in the previous section are not applicable to these sites.

We assume that there are total $N$ sites in the unit cell.
Within the vortex-free sector,
the Hamiltonian of our system under an applied magnetic field with $h_xh_yh_z\neq 0$ is expressed as,
\begin{eqnarray}
\mathcal{H} &=& \mathcal{H}_K + \mathcal{H}^{(3)} - \sum_{l}(h_{z}S^{z}_{l,1} + h_{z}S^{z}_{l,N}),\label{eq:totalham}\\
&=:& \frac{1}{2}\sum_{(l,m),(l^{\prime},n)} c_{l,m}A_{(l,m) (l^{\prime},n)}c_{l^{\prime},n}, \label{eq:totalhamA}\\
\mathcal{H}_K&=&-\sum_{ \Braket{ij}_{\alpha}} K_{\alpha}S_i^{\alpha}S_j^{\alpha},  \label{eq:ham1}\\
\mathcal{H}^{(3)} &=& -\Delta\sum_{i,j,k} S_i^{x}S_j^{y}S_k^{z},\label{eq:ham3}\\
\Delta &\sim& \frac{h_{x}h_{y}h_{z}}{K^{2}},
\end{eqnarray}
where $ \mathcal{H}_K$ is the Hamiltonian of the Kitaev model, $\mathcal{H}^{(3)}$ is the mass term generated by third order perturbations in the magnetic field, and the last term in the first line is the Zeeman term for edge spins. $\Braket{ij}_{\alpha}$, $\Braket{jk}_{\beta}$ with $\alpha \neq \beta$, and we assume $K = K_x = K_y = K_z$
here for simplicity in the derivation of $\Delta$. Details are included in
Supplemental Material (SM)~\cite{SM}.
Although the first-order term in the magnetic field is suppressed in the bulk of the spin liquid state, the edge Zeeman term is not negligible, 
as described below.
In the second line of Eq.~(\ref{eq:totalhamA}), $l$ and $l^\prime$ label the indices of the unit cell,
while $m$ and $n$ label the sites inside the unit cell.
A key idea of the derivation of the Majorana Hamiltonian Eq.~(\ref{eq:totalhamA}) is to identify the gauge-field Majorana operators $b^z_{l,1}$ and $b^z_{l,N}$  with new matter Majorana operators
$c_{l,0}$ and $c_{l,N+1}$, respectively,
and $m$ (and $n$) runs from 0 to $N+1$. This enables us to treat the edge Zeeman term exactly.
We write the coordinate of the $(l,m)$ site
as $\bm{r}_{l,m} = (x_{l,m}, y_{l,m})$.
We use the Fourier transformation of $c_{l,m}$
only in the $x$-direction; $c_{l,m} \rightarrow c_{k_x,m}$
Then, the Bloch Hamiltonian can be written as,
\begin{eqnarray}
\mathcal{H} &=& \frac{1}{2}\sum_{k_x,m,n}c_{k_{x},m}^\dagger\mathcal{H}_{k_{x},m,n}c_{k_{x},n}, \\
\mathcal{H}_{k_{x},m,n} &:=& 2\sum_{l, l^\prime}e^{-\textrm{i}k_{x}(x_{l,m}-x_{l^{\prime},n})}A_{(l,m) (l^{\prime},n)}.
\end{eqnarray}

{\it Definition of the energy current and the spin current.}---
To evaluate the spin Seebeck coefficient, we introduce the energy polarization operator as,
\begin{eqnarray}
\bm{P}_{E} = \frac{1}{2}\sum_{(l,m)(l^\prime,n)}\frac{\bm{r}_{l,m}+\bm{r}_{l^\prime,n}}{2} c_{l,m}A_{(l,m) (l^{\prime},n)}c_{l^{\prime},n},
\end{eqnarray} 
where $A_{(l,m) (l^{\prime},n)}$ is defined in {Eq.~(\ref{eq:totalhamA})}.
We introduce the energy current operator as $\bm{J}_{E} =\textrm{i} \comm{\mathcal{H}}{\bm{P}_{E}}$.
Note that since the chemical potential is always zero in the Kitaev system, $\bm{J}_{E}$ is equal to the thermal current operator.

We emphasize that although it is impossible to express the spin and spin current
operators in the bulk only in terms of $c$-operators,
the spin operator at the edge can be still written
solely in terms of matter Majorana fermions, because we redefined
$b^z_{l,1}$ and $b^z_{l,N}$ as $c_{l,0}$ and $c_{l,N+1}$, respectively.
The total $S^z$ at the edge, $S_z^\textrm{edge}=\sum_{l} \left[  S^{z}_{l,1}+S^{z}_{l,N}\right]$,
can be written in terms of a skew symmetric matrix $B_{(l,m)(l^\prime,n)}$,
\begin{eqnarray}
S_z^\textrm{edge}
=: \frac{1}{2}\sum_{(l,m),(l^{\prime},n)} c_{l,m}B_{(l,m)(l^\prime,n)}c_{l^\prime,n}\label{eq:totalhamB}.
\end{eqnarray}
We define the spin current operator using this $B$ matrix as,
\begin{eqnarray}
J_{s}^x := \frac{1}{4} \sum_{k_x,m,n}\left[v_{k_x}S^{z}_{k_x}+S^{z}_{k_x}v_{k_x}\right]_{mn}c^{\dagger}_{k_x,m}c_{k_x,n},
\end{eqnarray}
\begin{eqnarray}
[S^{z}_{k_x}]_{m,n} := \sum_{l,l'} 2e^{-\textrm{i}k_{x}(x_{l,m}-x_{l^{\prime},n})}B_{(l,m)(l^\prime,n)},
\end{eqnarray}
where $v_{k_x} = \frac{\partial \mathcal{H}_{k_x}}{\partial k_x}$.
There is an ambiguity in the definition of the spin current when spins are not conserved.
Thus, we simply employ a conventional definition relevant to experimental detection, using the anticommutation of the group velocity $v_{k_x}$ and the edge spin $S^{z}_{k_x}$.

\begin{figure}[b]
\includegraphics[width=8cm]{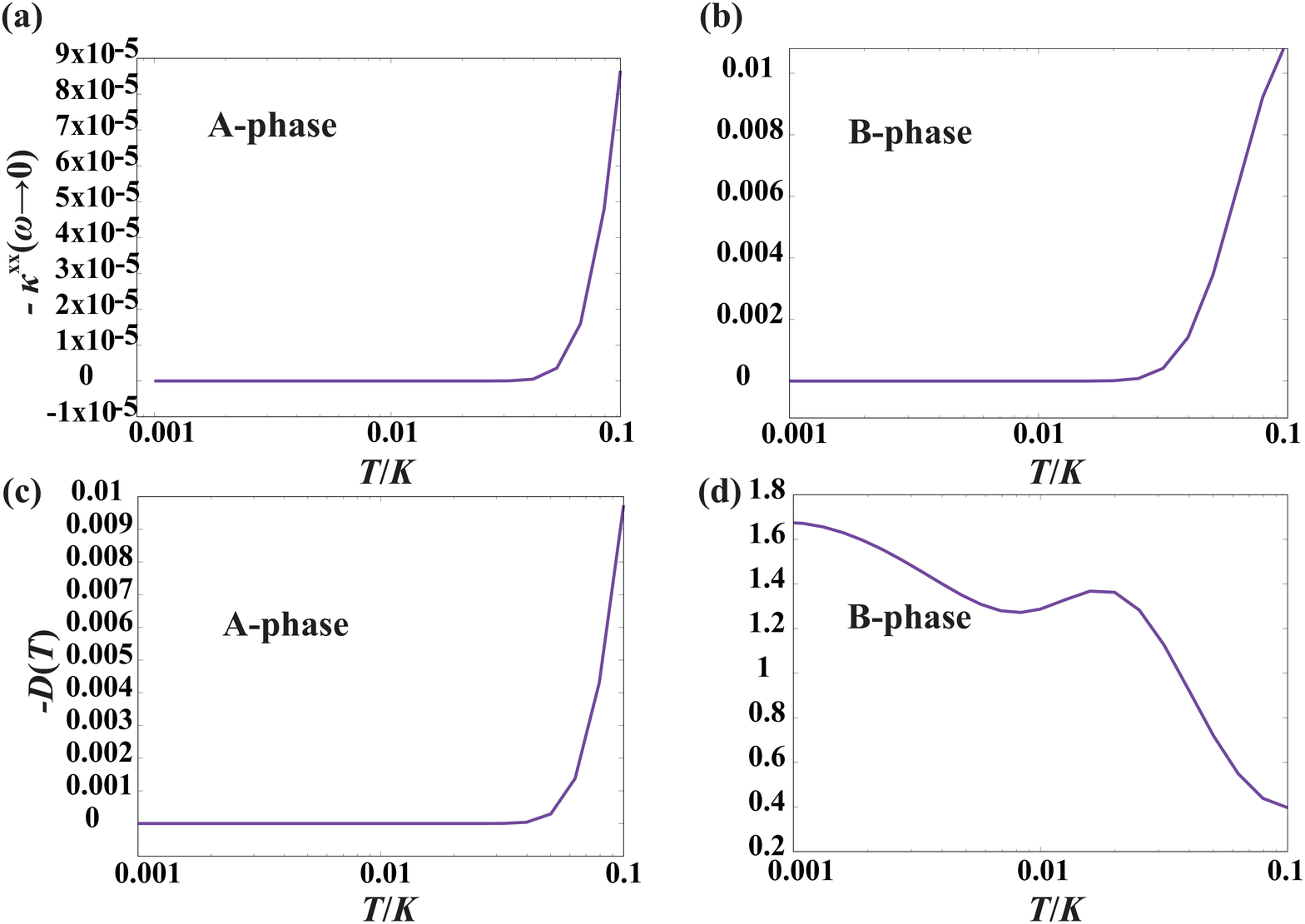}
\caption{ Plots of $\kappa^{reg}$ and $D(T)$ versus $T$. 
(a) $\kappa^{reg}$ for the $A$-phase. (b) $\kappa^{reg}$ for the $B$-phase. (c) $D(T)$ for the $A$-phase. (d) $D(T)$ for the $B$-phase. The parameters for the $A$-phase are $(K_x,K_y,K_z,h_z,\Delta,N,L_x)=(1.0,0.1,0.1,0.1,0,80,1000)$. The parameters for the $B$-phase are $(K_x,K_y,K_z,h_z,\Delta,N,L_x)=(1.0,1.0,1.0,0.1,0.01,80,1000)$.}
\label{fig:THC}
\end{figure}

{\it Edge spin Seebeck effect.}---
The spin Seebeck effect is characterized by the Kubo correlation function $\kappa^{xx}_\textrm{Kubo}$ for the spin current and the energy current as
$\Braket{J^{x}_{s}}_{\nabla T}/L_x = -\kappa^{xx}_\textrm{Kubo}\nabla_{x} T$.
We note that the longitudinal component is obtained directly from the Kubo formula, whereas the transverse component needs a contribution from the gravitational magnetization additionally~\cite{nasu2017thermal,nomura,sumiyoshi}.
We compute $\kappa^{xx}_\textrm{Kubo}$ as described in SM~\cite{SM},
and obtain,
\begin{eqnarray}
\kappa^{xx}_{\textrm{Kubo}}(T,\omega) &=& 2\pi D(T)\delta(\omega) + \kappa^{reg}(T,\omega),\label{eq:kappa1}\\
D(T) &=&-\frac{1}{TL_{x}}\sum_{k_{x},u,v,E_{k_{x},u}= E_{k_{x},v}} J^{E}_{k_{x},u,v}J^{s}_{k_{x},v,u} f'(E_{k_{x},u}), \label{eq:kappa2}\\
\kappa^{reg}(T,\omega) &=& -\frac{2\pi}{TL_{x}}\sum_{k_{x},u,v,E_{k_{x},u}\neq E_{k_{x},v}} J^{E}_{k_{x},u,v}J^{s}_{k_{x},v,u} \nn\\
&\times& \delta(\omega + E_{k_{x},v} - E_{k_{x},u}) \frac{f(E_{k_{x},v})-f(E_{k_{x},u})}{E_{k_{x},v}-E_{k_{x},u}},\label{eq:kappa3}
\end{eqnarray}
where the $D(T)$ is the Drude weight and $\kappa^{reg}(T,\omega)$ is the regular part of $\kappa_{\textrm{Kubo}}^{xx}(T,\omega)$, $E_{k_{x},u}$ is an eigenvalue of the Bloch Hamiltonian, and $f(E)$ is the Fermi distribution function.
$J^{s}_{k_{x},u,v}$ and $J^{E}_{k_{x},v,u}$ are, respectively, spin and energy currents. 

{\it Ballistic edge transport.}---
By using the Kubo formula obtained in the previous section, we calculate the spin Seebeck conductivity $\kappa^{xx}_\textrm{Kubo}$. We, here, present the numerical results for the Drude weight and the regular part of $\kappa_{\textrm{Kubo}}^{xx}(T,\omega)$ in the dc limit, \textit{i.e.} $\omega \rightarrow 0$.
We note that the regular part and the Drude Weight are always negative in our definition of the spin current.

To confirm whether the existence of edge states affects the spin Seebeck effect or not, we first perform the calculation for both the $A$-phase and the $B$-phase of the Kitaev model as shown in {Fig.~\ref{fig:THC}}.
The parameters for {Fig.~\ref{fig:THC}}{(a)} and {(c)} are $K_x=1$, $K_y=K_z=0.1$, $\Delta=0$, $h_z=0.1$, $N=80$, and $L_x=1000$, which correspond to the gapped $A$-phase.
The parameters for {Fig.~\ref{fig:THC}}{(b)} and {(d)} are $K_X=K_y=K_z=1$, $\Delta=0.01$, $h_z=0.1$, $N=80$, and $L_x=1000$, which correspond to the gapped $B$-phase with a chiral edge mode.
It is cautioned that the calculations are valid only for low temperature regions below 0.01 K, where the effective Hamiltonian is applicable. We show the results in high temperature regions above 0.01 K just for examining the contributions of the bulk gapful excitations to the regular part of the conductivity.
As shown in {Figs.~\ref{fig:THC}}{(a)}-{(b)}, $\kappa^{reg}$ has only the bulk contribution.
By comparing {Fig.~\ref{fig:THC}}{(c)} and {(d)} we can easily see that there is an edge contribution only in the $B$-phase.
The fact that the contribution of the chiral edge mode appears in the Drude weight means that the transport via the edge mode is protected from backscatterings, resulting in the generation of ballistic spin current at the edge.

\begin{figure}[b]
\includegraphics[width=8cm]{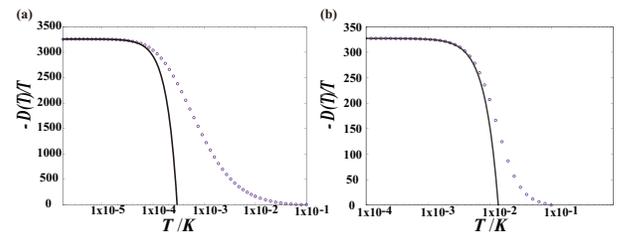}
\caption{Logarithmic plot of $D(T)/T$ versus $T/K$.
(a) For the parameter set $(h_z,\Delta,N,L_x) = (0.1,0.05,80,1000)$. White circles are the numerically calculated data, and the black line is the line fitting between $T=2.0 \times 10^{-6}$ and $T=2.0\times10^{-5}$. (b) $(h_z,\Delta,N,L_x) = (1.0,0.05,80,1000)$. The black line is the line fitting between $T=10^{-4}$ and $T=10^{-3}$.}
\label{fig:DW}
\end{figure}

An important feature of the Drude weight arising from
chiral Majorana edge contributions is the universal temperature scaling at low temperatures.
In {Fig.~\ref{fig:DW}}, we show the temperature dependence of the Drude weight divided by the temperature $D(T)/T$ at low temperature with $h_z=0.1,1.0$.
In the case of (a), the parameters are set to $K_x=K_y=K_z=1$, $\Delta=0.05$, $N=80$, and $L_x=1000$.
In the case of (a), the data are fitted by the fitting function shown in the solid line in the region from $T=2.0\times10^{-6}$ to $T=2.0\times10^{-5}$.
We use the fitting function $g(T) = a + bT^{c}$, and the result is that $a=3258.68(2)$, $b=-7.321(3)\times 10^{10}$, $c=2.08(2)$.
In the case of (b), the data are fitted by the fitting function shown in the solid line in the region from $T=10^{-4}$ to $T=10^{-3}$.
The result is that $a=327.027(3)$, $b=-1.9(2)\times 10^{6}$, $c=2.02(2)$.
This result shows nearly $T^2$ correction to $D(T)/T$ at low temperature.
We stress that this temperature dependence is robust against any perturbations due to disorder or phonons, because it arises from the chiral character of the Majorana edge states with no backscatterings.
This signature can be utilized for a clear-cut experimental detection of the Majorana edge states.

{\it Conformal field theory description.}---
To see the property of the dissipationless transport in more details, we further investigate the Drude weight from a different perspective.
This transport problem of the chiral edge mode is essentially one-dimensional (1D), and
conformal field theory (CFT) is known to be a powerful tool to investigate such a 1D system.

\begin{figure}[b]
\includegraphics[width=8cm]{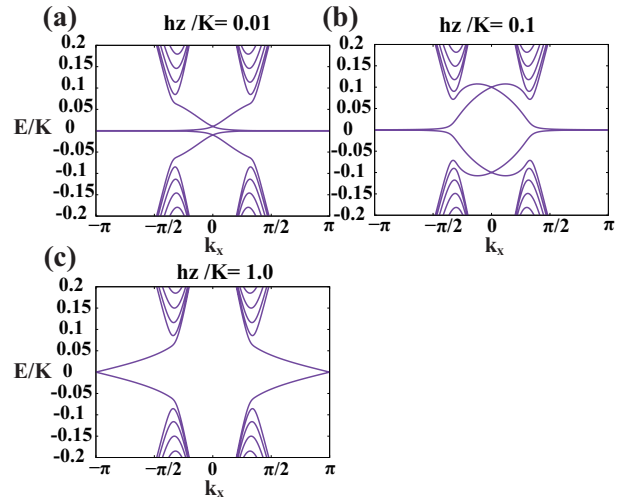}
\caption{ {Majorana band structures in a magnetic field.} ({a}) For the parameter set $(h_z,\Delta,N,L_x) = (0.01,0.05,80,1000)$. 
({b}) For the parameter set $(h_z,\Delta,N,L_x) = (0.1,0.05,80,1000)$. 
({c}) For the parameter set $(h_z,\Delta,N,L_x) = (1.0,0.05,80,1000)$. }
\label{fig:band}
\end{figure}

Before going into the CFT description, we show band structures (with gauge-field Majorana fermions at the edge) in {Fig.~\ref{fig:band}}.
In {Fig.~\ref{fig:band}}{(a)}-{(c)}, we change the magnetic field in the $z$-direction from $h_z=0.01$ to $h_z=1.0$.
The  magnetic field $h_z$ determines the coupling between matter Majorana fermions and gauge-field Majorana fermions. 
In the small coupling region shown in {Fig.~\ref{fig:band}}({a}), the cross point where the edge state passes zero energy is not well-defined because the energy band of the edge state becomes flat.
However, as we increase the coupling strength, the edge state becomes dispersive as shown in {Fig.~\ref{fig:band}}{(b)}-{(c)}.
In other words, the unique flat band of the zigzag edge with $h_z=0$ becomes chiral due to the mixing with gauge-field Majorana fermions, so that we can expect that CFT is applicable in the large-field region.
Thus, we investigate the temperature dependence of the Drude weight in the region where $h_z$ is large enough, and compare the results with the CFT prediction.
The Drude weight behaver at low temperature shown in {Fig.~\ref{fig:DW}}
is consistent with the Ising CFT when the magnetic field becomes large enough~\cite{Caselle_2002,fujimoto2003}.
From the prediction of CFT, the correction of $D(T)$ from the $T$-linear contribution always begins from $T^2$.
It is known that the leading irrelevant operator of the chiral Ising CFT is the energy-momentum tensor to the second power~\cite{Caselle_2002}, which leads to the $T^2$ correction to $D(T)/T$.

{\it Discussion and conclusion.}---
From the results obtained above, we can propose an experiment which potentially detects Majorana fermions in $\alpha$-RuCl$_3$, which is currently still under debate.
The dissipationless spin current generated by the spin Zeebeck effect is a unique property of the Kitaev spin liquid
which possesses chiral Majorana edge states.
The universal temperature scaling of the Drude weight in the spin Seebeck effect should be a definite signature of chiral Majorana edge states, and 
better observable, because we can expect that the universal scaling is stable with respect to various extrinsic perturbations such as disorder and phonons, provided that the roughness of the edge is sufficiently weak.
In the case that the edge is a strongly random admixture of a zigzag type and an armchair type,
the spin current is suppressed.
 However, it is expected that, even in such situations, the spin current does not vanish completely as long as a magnetic field h$_z$ is applied.
 Although disorder due to impurities in real systems may broaden the Drude peak, the total weight is not affected because of the chiral character of the edge states. 
These results are contrasted to the case without a bulk energy gap, where 
the Drude weight is substantially suppressed by weak randomness~\cite{kao2021}.
Not only for the basic research, the discovery of the dissipationless spin transport should be the key to the direct application of the Kitaev spin liquid to 
spintronics technology.
We note that, as seen in Fig. \ref{fig:THC}(d), even at very low temperatures $T/K \sim 0.001$, 
the magnitude of the Drude weight part of the spin Seebeck conductivity is roughly 
$\sim Ka\mu_{\rm B}/\hbar$ with $K$ the Kitaev interaction, $a$ a lattice constant, $\mu_{\rm B}$ the Bohr magneton.
For $\alpha$-RuCl$_3$~\cite{RuCl3_exp1}, $\sim Ka\mu_{\rm B}/\hbar \sim 1000\mu_{\rm B}$ ($m^{-1}\cdot s^{-1}\times \mu_{\rm B}$), 
which implies that the effect is much enhanced compared to conventional thermoelectric effects of electrons in semiconductors. 
The origin of the notable enhancement is attributed two factors; one is the absence of the backward scattering in the chiral edge state, and the other one is the flatness of the edge Majorana band (see Fig. \ref{fig:band}), i.e. the large energy-derivative of the density of states, which arises from the unpaired gauge-field Majorana fields at the edges. We stress that these factors are ubiquitous features of the Kitaev's chiral spin liquid state.

We would like to thank Y. Matsuda, T. Morimoto, and T. Shibauchi for fruitful discussions.
This work was supported by JST CREST Grant No. JPMJCR19T5, Japan, and JSPS KAKENHI Grant No. JP21H01039.
D.T. is supported by a JSPS Fellowship for Young Scientists and by JSPS KAKENHI Grant No. JP20J20385.

\bibliography{scibib}
\bibliographystyle{apsrev}

\end{document}


\title{Supplemental Material for ``Dissipationless Spin Current Generation in Kitaev's Chiral Spin Liquid''}

\author{Daichi Takikawa}
\email{takikawa@blade.mp.es.osaka-u.ac.jp}
\affiliation{Department of Materials Engineering Science, Osaka University, Toyonaka 560-8531, Japan}
\author{Masahiko G. Yamada}
\affiliation{Department of Materials Engineering Science, Osaka University, Toyonaka 560-8531, Japan}
\author{Satoshi Fujimoto}
\affiliation{Department of Materials Engineering Science, Osaka University, Toyonaka 560-8531, Japan}

\date{\today}

\maketitle

\appendix
\section*{Kitaev Hamiltonian}
In this section, to establish notations, we briefly summarize basics of the Kitaev honeycomb model.
We start with the following 
Hamiltonian on the honeycomb lattice shown in {Fig.1}{(a)} of our main text,
\begin{eqnarray}
\mathcal{H}_K=-\sum_{ \Braket{ij}_{\alpha}} K_{\alpha}S_i^{\alpha}S_j^{\alpha},  \label{eq:ham1}
\end{eqnarray} 
where $S^{\alpha}_i$ is an $\alpha=x$, $y$, $z$ component of an $S=1/2$ spin operator at the $i$th site.
Here, $\Braket{ij}_{\alpha}$ denotes that the $i$th site and the $j$th site are the nearest-neighbor sites connected by an $\alpha$-bond on the honeycomb lattice.
$\alpha$-bonds ($\alpha=x$, $y$, $z$) are defined as shown in  {Fig.1}{(a)} of our main text.

The ideal Kitaev Hamiltonian $\mathcal{H}_K$ is exactly solvable in terms of the Majorana fermion representation:
\begin{eqnarray}
S^x_j=\frac{\textrm i}{2}b^x_jc_j, ~S^y_j=\frac{\textrm i}{2}b^y_jc_j, ~S^z_j=\frac{\textrm i}{2}b^z_jc_j,
\end{eqnarray}
where $b_j^{\alpha}$ ($\alpha=x$, $y$, $z$) and $c_j$ are Majorana fermion operators, and the Hilbert space where these operators act is restricted to satisfy $D_i \ket{\phi}=\ket{\phi}$ with $D_i=b_i^xb_i^yb_i^zc_i$.
In terms of the Majorana representation, $\mathcal{H}_K$ is expressed as
\begin{eqnarray}
\mathcal{H}_K=\frac{{\textrm i}}{4}\sum_{i,j}\hat{A}_{ij}c_ic_j,   \label{eq:HK}
\end{eqnarray}
where $\hat{A}_{ij}=K^{\alpha}\hat{u}^{\alpha}_{ij}/2$ and $\hat{u}^{\alpha}_{ij}={\textrm i}b_i^{\alpha}b_j^{\alpha}$ with $\Braket{ij}_{\alpha}$,
and otherwise $\hat{A}_{ij}=0$.
The $Z_2$ gauge fields $\hat{u}^{\alpha}_{ij}$ commute with $\mathcal{H}_K$, and can be replaced by eigenvalues $\pm 1$.

For $K^{\alpha}>0$, in the ground state, we can put $\hat{u}^{\alpha}_{ij} \rightarrow 1$, and hence, $\hat{A}^{\alpha}_{ij} \rightarrow K^{\alpha}/2$.
We use this flux-free approximation throughout this paper.  It is a good approximation for the low-temperature region we are interested in.
Then, {Eq.~\eqref{eq:HK}} is reduced to the Hamiltonian of free massless Majorana fermions, which can be diagonalized in the momentum representation.
The phase diagram of this model is
shown in {Fig.1}{(b)} of our main text.
In {Fig.1}{(b)} of our main text
, $A$-phase is a gapped toric code phase, while $B$-phase is a gapless phase.

In the $B$-phase,
when a magnetic field $\vec{h} = (h_x,h_y,h_z)$ satisfying $h_{x}h_{y}h_{z}\neq0$ is applied to the system, the Zeeman interaction,
\begin{eqnarray}
\mathcal{H}_{2}=-\sum_{i} (h_{x}S_i^{x}+h_{y}S_i^{y}+h_{z}S_i^{z})
\label{eq:ham2}
\end{eqnarray}
generates a mass gap for Majorana fermions in the bulk, and the system exhibits a chiral spin liquid state with
a chiral Majorana edge state.
The edge state is known to be described by the Ising conformal field theory (CFT)~\cite{kitaev2006}.
The mass term is obtained by a perturbative calculation up to the third order in $h_{\alpha}$, which leads to three-spin interaction terms,
\begin{eqnarray}
\mathcal{H}^{(3)} &=& -\Delta\sum_{i,j,k} S_i^{x}S_j^{y}S_k^{z},\label{eq:ham3}\\
\Delta &\sim& \frac{h_{x}h_{y}h_{z}}{K^{2}},
\end{eqnarray}
where $\Braket{ij}_{\alpha}$, $\Braket{jk}_{\beta}$ with $\alpha \neq \beta$, and we assume $K = K_x = K_y = K_z$
here for simplicity in the derivation of $\Delta$.
In terms of Majorana fields, $\mathcal{H}^{(3)}$ is written as,
\begin{eqnarray}
\mathcal{H}^{(3)} &=& \textrm{i}\frac{\Delta}{8}
\sum_{\langle\!\langle ij \rangle\!\rangle} c_{i}c_{j},\label{eq:ham4}
\end{eqnarray}
where $\langle\!\langle ij \rangle\!\rangle$ means
a next-nearest-neighbor hopping.  The direction of
the hopping is shown in {Fig.1}{(a)} of our main text.
This term yields the Majorana mass gap $\Delta_{Majorana} \sim \Delta$.
It is noted that the mass gap term also arises from the perturbation in the non-Kitaev interaction $\Gamma'$~\cite{takikawa2019,takikawa2020}.We also note that the second-order corrections in the magnetic field merely renormalize the magnitude of the Kitaev interaction, and do not affect the effective Hamiltonian qualitatively.

We stress, here, that in the vortex-free spin liquid state, the first-order corrections with respect to the Zeeman term {Eq.~\eqref{eq:ham2}} vanish, and the gauge-field Majorana fermions $b^{\alpha}_j$ can be completely eliminated
in the effective low-energy Hamiltonian in the bulk, i.e.
the system can be described only in terms of matter Majorana fermions $c_j$.
However, the situation drastically changes at open edges
as described below, which is a key factor of disspatinless spin currents at the edges.
\subsection*{Derivation of the conductivity}
Here, we show the detail of the derivation of the spin Seebeck conductivity {Eqs.~(12)-(14)} in the main text.
We first diagonalize the Hamiltonian, whose labels run from $0$ to $N+1$ because the unpaired gauge-field Majorana fermions at open edges can be regarded as additional itinerant Majorana fermions c$_{k_{x},0}$ and c$_{k_{x},N+1}$. After this identification, the system is equivalent to a free fermion system with open boundaries, and can be treated exactly.
\begin{eqnarray}
{\vec{c}_{k_x}} &=& (c_{k_{x},0}, c_{k_{x},1},\dots, c_{k_{x},N},c_{k_{x},N+1})^{T} \label{eq:supp1},\\
\mathcal{H} &=& \frac{1}{2}\sum_{k_x}\vec{c}_{k_x}^\dagger\mathcal{H}_{k_{x}}\vec{c}_{k_x},\\
\mathcal{H}_{k_{x},m,n} &:=& 2\sum_{l,l'}e^{-\textrm{i}k_{x}(x_{lm}-x_{l'n})}A_{(l,m)(l,n)}.\label{eq:supp2}
\end{eqnarray}

Due to the particle-hole symmetry, all eigenvalues appear in pair with opposite signs and the same absolute value.
Thus, we label positive eigenvalues $E_{k_{x},n}$ ($n = 1,2,\dots,\frac{N}{2},\frac{N}{2}+1$), and negative ones $E_{k_x,-n}$ ($=-E_{k_x,n}$).  
The Bloch Hamiltonian is diagonalized as,
\begin{eqnarray}
U_{k_x}^{\dagger}\mathcal{H}_{k_x}U_{k_x}&=& \textrm{diag}(E_{k_x,-\frac{N}{2}-1},E_{k_x,-\frac{N}{2}},\dots,E_{k_x,-1},E_{k_x,1},\nn\\
&&\dots, E_{k_x,\frac{N}{2}},E_{k_x,\frac{N}{2}+1}),
\end{eqnarray}
with a unitary matrix $U_{k_x}$.
We introduce a fermionic operator $f$ to diagonalize the Hamiltonian as, 
\begin{eqnarray}
\vec{f}_{k_x}^{\dagger} &=& \vec{c}_{k_x}^{\dagger} U_{k_x} ,  \ {\vec{f}_{k_x}} =U_{k_x}^{\dagger} \vec{c}_{k_x}\label{eq:supp6}
\end{eqnarray}
Thus, the Hamiltonian can be recast into,
\begin{eqnarray}
\mathcal{H}&=& \frac{1}{2}\sum_{k_x}\vec{c}_{k_x}^\dagger U_{k_x}U_{k_x}^{\dagger}\mathcal{H}_{k}U_{k_x}U_{k_x}^{\dagger}{\vec{c}_{k_x}},\\
&=&\frac{1}{2}\sum_{k_x}E_{k_x,n}f_{k_x,n}^\dagger f_{k_x,n},
\end{eqnarray}

Next, we define an energy current operator for the honeycomb Kitaev model.
We first introduce an energy polarization operator as,
\begin{eqnarray}
P_{E}^x &=& \frac{1}{2}\sum_{(m,l)(n,l')}\frac{x_{m,l}+x_{n,l'}}{2} c_{m,l}A_{(m,l)(n,l')}c_{n,l'}\\
\end{eqnarray}
Then, we define an energy current as,
\begin{eqnarray}
J_{E}^x &=& \textrm{i} \comm{\mathcal{H}}{P_{E}^x}.
\end{eqnarray}
We introduce a group velocity as,
\begin{eqnarray}
v_{k_x,m,n} &=& \frac{\partial \mathcal{H}_{k_x,m,n}}{\partial k_x}\\
&=&(-2\textrm{i})\sum_{l,l'}(x_{l,m}-x_{l',n})A_{(l,m)(l',n)}e^{-\textrm{i}k_x(x_{l,m}-x_{l',n})}\label{eq:supp18},
\end{eqnarray}
Then, the energy current is written as,
\begin{eqnarray}
J_{E}^x &=& \frac{1}{4}\sum_{k_x,m,n} c_{k_x,m}^\dagger[v_{k_x}\mathcal{H}_{k_x}+ \mathcal{H}_{k_x}v_{k_x}]_{m,n}c_{k_x,n}.
\end{eqnarray}
Using fermionic operators given by {Eq.~\eqref{eq:supp6}}, we can rewrite the energy current as,
\begin{eqnarray}
J_{E}^x &=& \sum_{k_x,u,v}J^{E}_{k_x,u,v}f^{\dagger}_{u,k_x}f_{v,k_x}, \label{eq:S15}\\
J^{E}_{k_x,u,v} &=& \frac{1}{4}(E_{u,k_x}+E_{v,k_x})\left[U_{k_x}^{\dagger}\frac{\partial \mathcal{H}_{k_x}}{\partial k_x}U_{k_x}\right]_{u,v}.
\end{eqnarray}
Similarly, we can rewrite the spin current operator as,
\begin{eqnarray}
J_{s}^x&=&\sum_{k_x,u,v} J^{s}_{k_x,u,v} f^{\dagger}_{u,k_x}f_{v,k_x}, \\
J^{s}_{k_x,u,v}&=&\frac{1}{4}\left[U_{k_x}^{\dagger}\left(\frac{\partial \mathcal{H}_{k_x}}{\partial k_x}S^{z}_{k_x}+S^{z}_{k_x}\frac{\partial \mathcal{H}_{k_x}}{\partial k_x}\right)U_{k_x}\right]_{u,v}.
\label{eq:S18}
\end{eqnarray}

Finally, the spin Seebeck conductivity $\kappa^{\mu \nu}$ ($\mu$, $\nu = x$, $y$) is defined by,
\begin{equation}
\langle J^{\mu}_{s} \rangle_{\nabla T}/ L_x = -\kappa^{\mu \nu}\nabla_{\nu} T,
\end{equation}
where $L_x$ is the number of unit cells and $\Braket{O}_{\nabla T}$ is the expectation value of $O$ in the presence of a thermal gradient $\nabla T$.
We here focus on the longitudinal component.
The longitudinal conductivity can be evaluated from the Kubo formula,
\begin{equation}
\kappa^{xx}_\textrm{Kubo}(\omega) = \frac{1}{TL_{x}}\int^{\infty}_{0}\textrm{d}t \ e^{\textrm{i}(\omega + \textrm{i}\delta)t}\int^{\beta}_{0}\textrm{d}\lambda \langle J^{x}_{E}(-\textrm{i}\lambda)J^{x}_{s}(t)\rangle,\label{eq:supp32}
\end{equation}
where $J^{x}_\alpha(t) = e^{{i} \mathcal{H}t} J^{x}_\alpha e^{-{i} \mathcal{H}t}$ ($\alpha=s$, $E$) and $\beta = 1/T$ is the inverse temperature.
Using {Eqs.~\eqref{eq:S15}-\eqref{eq:S18}}, the Wick's theorem, and $E_{k_x,n}=E_{-k_x,n}$ derived from the inversion symmetry, $\kappa^{xx}(\omega)$ can be recast into,
\begin{widetext}
\begin{eqnarray}
\kappa^{xx}_{\textrm{Kubo}}(T,\omega) &=& 2\pi D(T)\delta(\omega) + \kappa^{reg}(T,\omega),\label{eq:kappa1}\\
D(T) &=&-\frac{1}{TL_{x}}\sum_{k_{x},u,v,E_{k_{x},u}= E_{k_{x},v}} J^{E}_{k_{x},u,v}J^{s}_{k_{x},v,u} f'(E_{k_{x},u}), \label{eq:kappa2}\\
\kappa^{reg}(T,\omega) &=& -\frac{2\pi}{TL_{x}}\sum_{k_{x},u,v,E_{k_{x},u}\neq E_{k_{x},v}} J^{E}_{k_{x},u,v}J^{s}_{k_{x},v,u} \delta(\omega + E_{k_{x},v} - E_{k_{x},u})\frac{f(E_{k_{x},v})-f(E_{k_{x},u})}{E_{k_{x},v}-E_{k_{x},u}}. \label{eq:kappa3}
\end{eqnarray}
\end{widetext}
Here, we have used $J^{E}_{-k_x,-u,-v}=-J^{E}_{k_x,v,u}$ because $v_{-k_x} = -v_{k_x}$. 
$D(T)$ is the Drude weight of the spin Seebeck conductivity, which characterizes the ballistic transport at $\omega=0$. $f'(E)$ means the derivative of the Fermi distribution $f(E)=1/(e^{\beta E}+1)$.

\bibliography{scibib}
\bibliographystyle{apsrev}